# The Puzzling Detection of X-rays From Pluto by *Chandra*

C. M. Lisse[1], R. L. McNutt, Jr.[1], S. J. Wolk[2], F. Bagenal[3], S. A. Stern[4], G. R. Gladstone[5], T. E. Cravens[6], M. E. Hill[1], P. Kollmann[1], H. A. Weaver[1], D. F. Strobel[7], H. A. Elliott[5], D. J. McComas[8], R.P. Binzel[9], B.T. Snios[10], A. Bhardwaj[11], A. Chutjian[12], L. A. Young[4], C.B. Olkin[4], and K.A. Ennico[13]

[1]  Space Exploration Sector, Johns Hopkins University Applied Physics Laboratory, 11100 Johns Hopkins Rd, Laurel, MD USA 20723   carey.lisse@jhuapl.edu, ralph.mcnutt@jhuapl.edu, matthew.hill@jhuapl.edu, peter.kollmann@jhuapl.edu, hal.weaver@jhuapl.edu

[2]  *Chandra* X-ray Center, Harvard-Smithsonian Center for Astrophysics, 60 Garden Street, Cambridge, MA, USA 02138 swolk@cfa.harvard.edu

[3]  Laboratory for Atmospheric and Space Physics, University of Colorado, Boulder, CO, USA 80303 bagenal@lasp.colorado.edu

[4]  Southwest Research Institute, Boulder, CO, USA 80302 alan@boulder.swri.edu, layoung@boulder.swri.edu, colkin@boulder.swri.edu

[5]  Southwest Research Institute, San Antonio, TX, USA 28510  randy.gladstone@swri.org,  helliott@swri.edu

[6]  University of Kansas, Dept. of Physics and Astronomy, Malott Hall, 1251 Wescoe Hall Dr., Lawrence, KS, USA 66045  cravens@ku.edu

[7]  Department of Earth and Planetary Sciences, Johns Hopkins University, Baltimore, MD, USA 21218 strobel@jhu.edu

[8] Dept. of Astrophysical Sciences, Princeton University, 4 Ivy Lane, Princeton, NJ 08544  dmccomas@princeton.edu

[9]  Dept. of Earth, Atmospheric, and Planetary Sciences, Massachusetts Institute of Technology, Cambridge, MA, USA 02139 rpb@mit.edu

[10]  Dept. of Physics, University of Connecticut, Storrs CT, USA 06269  snios@phys.uconn.edu

[11]  Space Physics Laboratory, Vikram Sarabhai Space Centre, Trivandrum 695022, INDIA Anil_Bhardwaj@vssc.gov.in, Bhardwaj_SPL@yahoo.com

[12]  Astrophysics and Space Sciences Section, Jet Propulsion Laboratory/Caltech, Pasadena CA 91109, USA ara.chutjian@jpl.nasa.gov

[13]  Astrophysics Branch, Space Sciences Division, NASA/Ames Research Center, Moffett Field, CA, USA 94035 Kimberly.Ennico@nasa.gov

17 Pages, 3 Figures, 2 Tables

Proposed Running Title: **PUZZLING X-RAY DETECTION OF PLUTO**

**Please address all future correspondence, reviews, proofs, etc. to:**

Dr. Carey M. Lisse

Planetary Exploration Group, Space Exploration Sector

Johns Hopkins University, Applied Physics Laboratory

11100 Johns Hopkins Rd

Laurel, MD 20723

240-228-0535 (office) / 240-228-8939 (fax)

Carey.Lisse@jhuapl.edu

# Abstract

Using *Chandra ACIS-S*, we have obtained low-resolution imaging X-ray spectrophotometry of the Pluto system in support of the New Horizons flyby on 14 July 2015. Observations were obtained in a trial "seed" campaign conducted in one visit on 24 Feb 2014, and a follow-up campaign conducted soon after the New Horizons flyby that consisted of 3 visits spanning 26 Jul – 03 Aug 2015. In a total of 174 ksec of on-target time, in the $0.31 - 0.60$ keV passband, we measured 8 total photons in a co-moving 11 x 11 pixel$^2$ box (the 90% flux aperture determined by observations of fixed background sources in the field) measuring ~121,000 x 121,000 km$^2$ (or ~100 x 100 $R_{Pluto}$) at Pluto. No photons were detected from $0.60 - 1.0$ keV in this box during the same exposures. Allowing for background, we find a net signal of 6.8 counts and a statistical noise level of 1.2 counts, for a detection of Pluto in this passband at > 99.95% confidence. The Pluto photons do not have the spectral shape of the background, are coincident with a 90% flux aperture co-moving with Pluto, and are not confused with any background source, so we consider them as sourced from the Pluto system. The mean 0.31 - 0.60 keV X-ray power from Pluto is 200 $^{+200/}_{-100}$ MW, in the middle range of X-ray power levels seen for other known solar system emission sources: auroral precipitation, solar X-ray scattering, and charge exchange (CXE) between solar wind (SW) ions and atmospheric neutrals. We eliminate auroral effects as a source, as Pluto has no known magnetic field and the New Horizons *Alice* UV spectrometer detected no airglow from Pluto during the flyby. Nano-scale atmospheric haze particles could lead to enhanced resonant scattering of solar X-rays from Pluto, but the energy signature of the detected photons does not match the solar spectrum and estimates of Pluto's scattered X-ray emission are 2 to 3 orders of magnitude below the $3.9 \pm 0.7$ x $10^{-5}$ cps found in our observations. Charge-exchange-driven emission from hydrogenic and heliogenic SW carbon, nitrogen, and oxygen (CNO) ions can produce the energy signature seen, and the 6 x $10^{25}$ neutral gas escape rate from Pluto deduced from New Horizons' data (Gladstone *et al.* 2016) can support the ~3.0 $^{+3.0/}_{-1.5}$ x $10^{24}$ X-ray photons/s emission rate required by our observations. Using the solar wind proton density and speed measured by the Solar Wind Around Pluto (SWAP) instrument in the vicinity of Pluto at the time of the photon emissions, we find a factor of 40 $^{+40/}_{-20}$ lower SW minor ions flowing planarly into an 11 x 11 pixel$^2$, 90% flux box centered on Pluto than are needed to support the observed emission rate. Hence, the SW must be somehow significantly focused and enhanced within 60,000 km (projected) of Pluto for this mechanism to work.

## Introduction

Pluto, the first and largest discovered Kuiper Belt Object, lies at the outer edges of our solar system and was the target of the 14 July 2015 flyby by the NASA New Horizons (NH) mission (Stern *et al.* 2015). Pluto is known to have an atmosphere which changes size and density with its seasons (Elliot *et al.* 1989, 2003, McNutt *et al.* 1989, Strobel *et al.* 2003). Preliminary simulation results of its atmosphere from the flyby revealed a majority $N_2$ atmosphere with a condensed exobase of ~1000 km height and a low escape rate of $< 7x10^{25}$ mol/sec (Stern *et al.* 2015, Gladstone *et al.* 2016). Pluto is also immersed in the interplanetary solar wind (SW), and how it interacts with the wind depends on the state of its atmosphere. This physical situation is similar to that of Mars in the SW at 1.5 Astronomical Units (AU) from the Sun, although the presence of a long extended plasma tail streaming downstream from Pluto (McComas *et al.* 2016) may have aspects of the comet case at 1 AU (Bodewits *et al.* 2007; Christian *et al.* 2010; Lisse *et al.* 1996, 2001, 2005, 2007, 2013; Wolk *et al.* 2009).

Given that most pre-encounter models of Pluto's atmosphere had predicted it to be much more extended, with an estimated loss rate to space of $\sim 10^{27}$ to $10^{28}$ mol/sec of $N_2$ and $CH_4$ (similar to the typical $H_2O$ loss rates for Jupiter Family Comets (JFC) comets at 1 AU), we attempted to detect X-ray emission created by SW-neutral gas charge exchange interactions in the low density neutral gas surrounding Pluto. Even though the solar illumination and the SW flux both decrease as $1/r^2$, causing a near-Pluto neutral molecule's lifetime against photionization and charge exchange to be measured in years rather than days (Bagenal *et al.* 2015), the projected *Chandra* pixel size increases as $r^2$. Hence, roughly the same number of total emitting X-ray centers should be in each *Chandra* projected 12,000 x 12,000 $km^2$ pixel for Pluto, as for a "typical" JFC comet observed by *Chandra* at 1 AU (e.g., 2P/Encke observed by *Chandra* in 2003 (Lisse *et al.* 2005) or 9P/Tempel 1 observed by *Chandra* in 2005 (Lisse *et al.* 2007)). Based on our previous JFC comet X-ray detections, and an estimated neutral gas escape rate $Q_{gas} \sim 3 \times 10^{27}$ mol/sec, we expected a total *Chandra* count rate for Pluto on the order of 3 x $10^{-5}$ cps. With an estimated chip background rate of ~1 x $10^{-6}$ cps, the major concern with observing Pluto was that any sky or particle backgrounds could dominate the observed X-ray signal.

In late 2013 we received 35 ksec (~10 hours) of *Chandra* time to image the system spectrophotometrically. Given the *Chandra* visibility window constraints for the Pluto system, the first observations were possible starting mid-February 2014. To maximize the potential signal from the *Chandra* observations, we worked to schedule the *Chandra* Pluto observations at a time when the variable SW flux as extrapolated from New Horizons to Pluto's location would be near its maximum. We used the SW trends measured by the NH Solar Wind Around Pluto (SWAP) instrument (McComas *et al.* 2008), which was ~4 AU upstream of Pluto at the time of our observations and had been monitoring the SW for almost a year previously while NH was in its "hibernation mode." At the time of the observations we had received downloaded NH data only through Oct 2013, and the need to extrapolate the SW conditions forward in time to late February 2014 introduced significant uncertainties in the extrapolation.

## Results and Analysis

### 24 Feb 2014 Observations

Spectral imaging observations of the Pluto system using the *Chandra* Advanced CCD Imaging Spectrometer (ACIS) - S-array (ACIS-S) were obtained under *Chandra* program #15699 using a single telescope sky pointing from 24 Feb 2014 02:02:51 to 12:17:15 UTC (Tables 1 and 2). The Pluto system was centered near the "sweet spot" of the *Chandra* S3 chip, where the instrument spectral imaging response is best behaved. *Chandra* did not track the motion of Pluto on the sky, but instead tracked sky-fixed targets at the nominal sidereal rate. The instrument was operated in Very Faint (VF) event-detection mode, and a total of 8700 counts were detected on the S3 chip during 35 ksec of observing. By filtering the detected events in energy (0.31 – 0.60 keV for charge exchange, and 0.8 – 2.0 keV for stellar photosphere emission), we found that we best removed the instrumental background signal while preserving the flux from astronomical sources. (The carbon K-shell edge is at 280 eV, close to the *Chandra* background peak spanning $250 \pm 50$ eV, making Pluto CV photons hard to distinguish from ACIS-S background photons. For this reason we have chosen to exclude X-ray photons below 310 eV range in this study.) Even after energy filtering, a low level of background counts was found throughout the *Chandra* field of view (FOV). The average number of counts per pixel across the array was < 1, necessitating signal analysis using small-number, Poisson statistics. Smoothing out the background using a very large, 30 x 30 pixel$^2$ Gaussian footprint produced a map which shows structure across the array similar to that expected from Röntgensatellit (ROSAT) 1/4 and 3/4 keV maps of the sky around (R.A.=283.60$^o$, DEC=-20.15$^o$). This argues that the dominant low energy X-ray background contribution in the data is from the sky background, consistent with other studies (Slavin *et al.* 2013, Wargelin *et al.* 2014).

As the ACIS-S3 FOV was tracking the sky at sidereal rates, stellar objects were fixed in pixel position, while Pluto slowly moved, at a rate of ~3.8" (or 7.5 ACIS-S pixels) per hour, with a total track length of ~72 pixels during our observations. Creating images of our data in sky-centered and Pluto-centered coordinates, we distinguished a number of stellar sources from the background. While the list of detected sources is only a small subset of the stars known to be in the field (Pluto was within 7$^o$ of the galactic plane on 24 February 2014), enough (6) were detected to register the field and determine the effective beamwidth during the observations. Using these we found that a 5.5-pixel-radius circle contains > 90% of the point sources flux for objects registering 10 to 100 counts total. Taking the 90% footprint and placing it over the nominal location of Pluto in the Plutocentric *Chandra* image, we found a total of 2.0 cts in the $0.3 - 0.6$ keV energy range. Placing the same footprint at 1000 locations gridded around Pluto in the same image, we found an average of $0.221 \pm 0.453$ (1$\sigma$) background cts. We could thus claim a net Pluto signal of $1.779 \pm 0.453$ (1$\sigma$) cts from the Feb 2014 *Chandra* observations, marginally significant above zero. From this marginal detection, and using previous *Chandra* observations of JFC comets (Bodewits *et al.* 2007; Christian *et al.* 2010; Lisse *et al.* 1996, 2001, 2005, 2007, 2013; Wolk *et al.* 2009) for calibration, NH SWAP's measurement of the SW flux (McComas *et al.* 2016), and the value of 33.2 AU for the Pluto-*Chandra* distance on 24 Feb 2014 we related this "detection" to the product of the SW flux and neutral gas production rate from Pluto and found $Q_{gas} \leq 1.5$ x $10^{28}$ mol/sec (3$\sigma$) (Lisse *et al.* 2015). This upper limit, assuming Pluto's atmospheric density fell off as $1/r^2$ like a comet's, was useful, in that it was consistent with the pre-encounter estimated $Q_{gas}$ rates of 2 x $10^{27}$ and 5 x $10^{27}$ mol/sec produced by global atmospheric models of Pluto (Tucker *et al.* 2015, Zhu *et al.* 2014).

**26 Jul – 03 Aug 2015 Measurements**

Using the positive results of these 35 ksec "seed" observations, we contacted the *Chandra* project and requested additional observing time during the New Horizons Pluto encounter. We were generously awarded another 145 ksec of observatory time to study Pluto using the same methodology and the NH *in situ* measurement of the Plutonian SW to determine robustly if our marginal detection was real. Due to *Chandra* pointing restrictions, we could not begin observing until 26 July 2015, almost 2 weeks after the 14 Jul 2015 Pluto encounter but were then able to integrate on-target for ~139 ksec in 3 visits over the timespan 26 Jul – 03 Aug 2015 (Table 1) in OBSIDs 17703, 17708, and 17709 (note that 6 ksec of off-target observatory overhead time was used to slew, point, and settle *Chandra* during these measurements). In all we detected another 6 counts at 0.3–0.6 keV, on top of a background of $1.20 \pm 1.16$ cts (Table 2). **We then** directly **co-added the three new 2015 integrations with the 2014 results by stacking Plutocentric image pixels registered on Pluto. In the combined total Plutocentric exposure, we find a total Pluto X-ray signal of $6.79 \pm 1.16$ cts, a total count rate of $3.9 \pm 0.7 \times 10^{-5}$ cps, and a probability that the 7 photons detected at Pluto due to a random fluctuation of the background as $< 5 \times 10^{-4}$** (assuming small number background statistics modeled by a Poisson distribution with $\lambda = 1.2$ cts/174 ksec = 0.95 cts/139 ksec) in an 11 x 11 pixel (5.5" x 5.5" or 121,000 x 121,000 km²) box centered on its ephemeris position (Figure 1).

At this point we performed one more careful check on the Pluto Chandra signal, and examined the ~30 stars in the Chandra FOV that were brighter than the [V]~14, [J]~13 for signs of an x-ray signal in case any red leak in the ACIS blocking filter was causing a false detection of Pluto. While we detected 13 objects in the 1.0 - 2.0 keV passband, none of these objects were detected in the 0.31 - 0.60 keV passband. We thus have high confidence that we have an actual detection of Pluto, as the Plutonian photons do not have the spectral shape of the background, are coincident with a 90% flux aperture co-moving with Pluto, are not confused with any background source, and are significant, i.e., not random background fluctuations, at the > 99.95 % confidence level (Figure 2).

## Discussion

Given this detection of 6.8 net (8 – 1.2 background) X-ray photons coming from the vicinity of Pluto, what could be their source? The 174 ksec of total *Chandra* ACIS-S on-target integration time required to detect these photons implies a significant total equivalent X-ray power of 200 $^{+200/}_{-100}$ MW, assuming a detected photon energy-weighted average Chandra ACIS-S effective area of 40 $^{+40/}_{-20}$ cm$^2$ in the Feb 2014 - Aug 2015 timeframe and a mean Chandra Pluto distance of 33 AU. (Note that the effective area is a strong function of both energy and time of observation in the 0.31 - 0.60 keV energy range - see http://cxc.harvard.edu/cgi-bin/prop_viewer/build_viewer.cgi ?ea). X-ray photons observed from sources in the solar system are typically produced by (1) precipitation of solar wind or magnetospheric energetic ions into planetary atmospheres, (2) fluorescent scattering of solar X-rays (Bhardwaj *et al.* 2007, Branduardi-Raymont *et al.* 2010, Collier *et al.* 2014, Cravens *et al.* 2003, Dennerl 2002, Dennerl *et al.* 2002, 2010), or (3) charge exchange of energetic solar wind or magnetospheric energetic ions with cometary or planetary neutrals.

However, no single one of these physical explanations is satisfactory for Pluto:

(1) It is unlikely for Pluto to have a significant intrinsic magnetic field (Bagenal *et al.* 1997) and McComas *et al.* (2016) sets an upper bound of < 30 nT at the planet's surface. Further, no auroral emissions were detected at Pluto using the Alice UV spectrograph during the New Horizons flyby on 14 July 2015 (Gladstone *et al.* 2016).

(2) We can rule out coherent scattering, as the detected energy spectrum of 8 X-rays in the 0.33 – 0.60 keV region and none in the 0.60 – 1.0 keV region is counter to the maximum count rate at ~1 keV we would expect for a solar X-ray spectrum convolved with the ACIS-S effective area function (Snios *et al.* 2014). The 0.3 – 0.6 keV photons are in the proper energy range to be due to resonant fluorescent K-shell scattering by N, C and O atoms in the $N_2$, $CH_4$, and CO ices on the surface of Pluto and the O atoms in the $H_2O$ ice on the surface of Charon, and the ratio of 5:1 C+N:O energy photons seems to agree with the ~4:1 ratio of Pluto:Charon's surface area. However, extrapolating Dennerl's *Chandra* ACIS-I observations of Martian solar x-ray scattering at 1.5 AU to Pluto at 33 AU produces count rates ~3 orders of magnitude lower than measured, assuming that the bulk of the Martian x-rays are scattered at 100 - 150 km altitude above the Martian surface at ambient pressures similar to Pluto's surface pressure of 10 μbar (Dennerl 2002). A similar result is found by comparing Elsner *et al.* (2002)'s Europa x-ray emission measurements taken by Chandra to our Pluto results, suggesting it would be very hard to produce the observed x-rays via scattering from an icy surface. To match our measured intensity, an increase in either solar X-ray activity or the number of efficient x-ray scatterers would be required. However, extrapolation of measurements from the GOES X-ray sensor, an Earth-orbiting satellite that tracks solar X-ray emissions, indicate that the Pluto observations were taken during a period of quiet solar X-ray activity. Pluto does have layers of fine particulate haze suspended in its atmosphere, to at least 400 km above its surface (Gladstone *et al.* 2016, Cheng *et al.* 2017). A very highly concentrated, 10 - 1000 Å collection of haze grains composed of C, N, and O atoms fluorescing under the sun's X-ray insolation could produce significant resonant scattering, but it seems difficult to produce the power observed by Chandra at 0.60 to 1.37° phase (Table 1) with the relatively diffuse $\tau = 0.004$ normal optical depth haze found by New Horizons.

(3) X-ray emission via charge exchange between highly stripped hydrogenic and heliogenic minor ions in the solar wind and neutral gas species in comets and planetary atmospheres has been known to exist since its discovery by ROSAT observations of comet Hyakutake (Lisse *et al.* 1996) and has been detected from the short period JFC comet population for all objects within a few AU of the Sun with loss/escape rate $Q_{gas} > 1$ x $10^{27}$ mol/sec. Following the models of Cravens (1997), we expect the X-ray emission rate to trend linearly as the objects' $Q_{gas}$. Results from the NH Alice UV occultations and NH SWAP SW bowshock measurements for the neutral atmosphere escape rate are consistent with $Q_{gas} \sim 5$ x $10^{25}$ mol/sec, 50 times lower than pre-encounter estimates, and comprised of $CH_4$ instead of $N_2$ (Gladstone *et al.* 2016, Bagenal *et al.* 2016, Zhu *et al.* 2014). This outflow rate is capable of supporting the $\sim 3^{+3/}_{-1.5}$ x $10^{24}$ X-ray photons/sec emission rate required to produce the $3.9 \pm 0.7$ x $10^{-5}$ cps seen by a Chandra ACIS-S detector with $\sim 40$ cm$^2$ of effective collecting area if only 5% of the outflowing neutrals produce charge exchange X-rays. (McComas *et al.* (2016) reported a heavy ion tail immediately behind Pluto with a $CH_4^+$ flux $\sim 5$ x$10^{23}$ s$^{-1}$ from SWAP measurements.) The observed total $L_x$ of $200^{+200/}_{-100}$ MW for a $V_{mag} = 14$ object puts Pluto in the region of CXE driven emission for solar system X-ray sources (Figure 3). However, using the SW density ( $<n_{SW}>$ = 0.0115/cm$^3$) and speed ($<V_{SW}>$ = 376 km/sec) at the time of the Chandra observations determined by extrapolating New Horizons SWAP instrument proton measurements back to Pluto (Table 2), and a 1x$10^{-3}$ SW minor ion:proton abundance ratio (Schwadron & Cravens 2001), we find 7.6 $^{+7.6/}_{-3.8}$ x $10^{22}$/sec, or a factor of 40 $^{+40/}_{-20}$ , fewer SW minor ions than are needed in a $\sim 121,000$ x

121,000 km box to support a $\sim 3.0$ $^{+3.0/}$$_{-1.5}$ x $10^{24}$ X-ray photons/sec emission rate. However, while soft 0.31 - 0.60 keV energy range photons are favored by CXE, for most comets near the ecliptic plane oxygen minor ion photons in the range 500 - 600 eV dominate (i.e, the C+N:O photon ratio < 1), suggesting a rather different SW minor ion composition at Pluto than at 1 AU. (There is a suggestion of SW spatial composition variability: 2013 *Chandra* observations of comet C/PANSTARRS 2013 L4 at r = 1.1 AU and +84$^o$ heliographic latitude (Snios *et al.* 2016) found a low ionization polar wind that was C-rich and O-poor.) Either our assumptions of how the SW density and speed propagate from $\sim$1 AU out to 33 AU are incorrect, or the SW must be somehow significantly focused and enhanced, and its composition altered within the 60,000 km (50 $R_{Pluto}$) projected radius from Pluto contained in our 11 x 11 pixel$^2$ Chandra box, in an interaction very different than that of the comet-SW case for this mechanism to produce the photons seen. Further, the $L_x = 2.0^{+2.0/}$$_{-1.0}$ x $10^{15}$ erg/sec (200 MW) of X-ray luminosity found for Pluto at $\sim$33 AU in this work for $Q_{gas} = 5$ x $10^{25}$/sec is about 5 times the $L_x = 3.8$ x $10^{14}$ erg/sec found for comet 2P/Encke with $Q_{gas} = 2$ x $10^{27}$/sec at $r_h = 0.88$ AU in November 2003 using Chandra ACI-S (Lisse *et al.* 2005). For CXE to be the source of the emission Pluto would need to be at least 150 times more efficient per outflowing neutral gas molecule at making x-rays than comets. Given that a neutral gas molecule lasts for years after release from Pluto, rather than 1-10 days after release from a comet at 1 AU, this is quite feasible.

We are thus currently left with no obvious (i.e., known from observations of other solar system X-ray sources) mechanism for producing the measured X-rays detected from the vicinity of Pluto by *Chandra*. As the photon detections seem robust, in the absence of any clear working hypothesis, we need to speculate about possible new mechanisms. Could resonant scattering by abundant nm-sized organic haze grains in Pluto's atmosphere (Gladstone *et al.* 2016, Stern *et al.* 2015) greatly enhance its X-ray backscattering efficiency and be the cause of Pluto's high observed X-ray count rate? Or could draping of the solar wind and interplanetary magnetic field be focusing more solar wind minor ions into the region around Pluto than expected, increasing the production of charge exchange X-rays, perhaps in the "tail" region downstream of Pluto from the Sun (Bagenal *et al.* 2016, McComas *et al.* 2016). Could this draping somehow affect the abundance ratios in the "tail", or is it possible that oxygen minor ions are preferentially removed (versus carbon and nitrogen ions) as the SW propagates out to 33 AU and beyond? Another possibility for the great X-ray production efficiency we see for Pluto is the long neutral gas lifetime vs. ionization at 33 AU. Gas released from Pluto at ~10 m/sec survives in a neutral state for years, as compared to the 1-10 days for gas released at ~500 m/sec by comets at 1 AU, while the SW speeds are comparable at ~400 km/sec, meaning that Pluto's neutrals flow 10 times farther before ionizing and thus should have time to encounter ~100 times more SW minor ions. I.e., the effective interaction region between Pluto's released neutrals and SW minor ions may be much larger than naively expected (e.g., if the long lifetime of the slowly moving neutrals at 33 AU leads to the formation of a neutral gas Pluto torus centered around Pluto's orbit).

## Summary and Conclusions

- The total signal measured from Pluto in an 11 pixel x 11 pixel box co-moving with Pluto is found to be 8 photons in 174 ksec from 0.31 – 0.60 keV. No photons from 0.60 – 1.0 keV are found in the same exposures.

- The net signal is 8 - 1.2 = 6.8 photons, including all backgrounds, instrumental and sky. The background levels measured in one thousand boxes measuring 11 pixel x 11 pixel spread across the chip appear to be normally distributed at 1.21 ± 1.16 counts. The confidence level (assuming Poisson statistics) of our Pluto detection is at the  > 99.95 % level, and the corresponding total X-ray luminosity $L_x = 2.0$ $^{+2.0}_{-1.0}$ x $10^{15}$ erg/sec (200 MW).

- There is no obvious background source confused with Pluto in the four different Chandra visits.

- We see no evidence for an extended signal beyond the central 90% point spread function (PSF) centered on Pluto.

- Six of the photons detected lie in the 370 - 470 eV range of N VI CXE, and one of the photons lies in the N VII range. Alternatively, five of the photons are at or above the K-shell edge (resonance edge) of nitrogen (400 eV), and one of the photons exhibits an energy of ~ 596 eV, above the 530 eV oxygen K-shell edge.  The lowest three energy photons could be from C V emission instead of N VI, given the energy broadening (±50 eV) of the ACIS-S detector. OVII photons are rare from Pluto.

- The observed emission from Pluto is not aurorally driven. If due to scattering, it would have to be sourced by a unique population of nanoscale haze grains composed of C, N, and O atoms in Pluto's atmosphere resonantly fluorescing under the Sun's insolation. If driven by charge exchange between SW minor ions and neutral gas species (mainly $CH_4$) escaping from Pluto, then density enhancement and adjustment of the SW minor ion relative abundance in the interaction region near Pluto is required versus naïve models.

## Acknowledgements

The authors would like to thank NASA for financial support of the New Horizons project, and we thank the entire New Horizons mission team for making the results of the flyby possible. The authors would also like to gratefully acknowledge the *Chandra* X-ray Center and the *Chandra* project, especially director Dr. Wilkes, for their generous and vital support of our work. C. Lisse would also like to acknowledge support for this project from *Chandra* GO grant CXC-GO3-15100843, and S. Wolk was supported by NASA contract NAS8-03060.

# Tables

## Table 1 - Observing Circumstances of the *Chandra* Pluto Observations[a]

| OBSID[b] | Start Time | RA[c] | Dec[c] | $r_{h,Pluto}$[d] | $\Delta_{Pluto}$[e] | $Phase_{Pluto}$[f] | $Elong_{Pluto}$[g] | $[V]_{Pluto}$ |
|---|---|---|---|---|---|---|---|---|
| | | | | (AU) | (AU) | (deg) | (deg) | (mag) |
| **15699** | **2014-02-24T02:02:51** | 18 54 21.39 | -20 09 05.7 | 32.6 | 33.2 | 1.37 | 52.4 | 14.2 |
| **17703** | **2015-07-26T23:50:00** | 18 58 00.34 | -20 48 45.6 | 32.9 | 32.0 | 0.60 | 160. | 14.1 |
| **17708** | **2015-07-30T05:11:59** | 18 57 37.51 | -20 49 47.8 | 32.9 | 32.0 | 0.71 | 156. | 14.1 |
| **17709** | **2015-08-01T21:11:12** | 18 57 26.32 | -20 50 19.8 | 32.9 | 32.0 | 0.76 | 154. | 14.1 |

[a] - As Seen From *Chandra*
[b] - *Chandra* OBserving Sequence IDentification number.
[c] – ACIS-S3 sweet spot placed at Pluto's location at the mid-point of the observation.
[d] - Pluto-Sun distance at the mid-point of the observation.
[e] - Pluto- Chandra distance at the mid-point of the observation.
[f] - Sun-Pluto-Chandra and Sun-Chandra-Pluto angles at the mid-point of the observation.

## Table 2 - Circumstances of the Detected Pluto Photons[a,b,c]

| OBSID | $I_{time}$[a] (ksec) | Total Chip Rate (cps) | Total On Chip # of 310 - 600 eV Photons | $V_{SW}$[d] (km sec$^{-1}$) | $n_{SW}$[d] (cm$^{-3}$) |
|---|---|---|---|---|---|
| **15699** | **35.1 ksec** | **0.262 cps** | **2816** | | |
| Pluto Photon Energy = 597 eV | Pluto Photon Time | = 2014-02-24T02:02:51 + 3.55 hrs | | 364.7 | 0.01230 |
| Pluto Photon Energy = 396 eV | Pluto Photon Time | = 2014-02-24T02:02:51 + 8.40 hrs | | 364.4 | 0.01249 |
| **17703** | **14.1 ksec** | **0.213 cps** | **1174** | | |
| Pluto Photon Energy = 374 eV | Pluto Photon Time | = 2015-07-26T23:50:00 + 3.26 hrs | | 393.0 | 0.01016 |
| **17708** | **17.6 ksec** | **0.215 cps** | **1146** | | |
| Pluto Photon Energy = 327 eV | Pluto Photon Time | = 2015-07-30T05:11:59 + 4.89 hrs | | 381.5 | 0.01286 |
| **17709** | **107.2 ksec** | **0.217 cps** | **9099** | | |
| Pluto Photon Energy = 465 eV | Pluto Photon Time | = 2015-08-01T21:11:12 + 0.81 hrs | | 374.7 | 0.01399 |
| Pluto Photon Energy = 405 eV | Pluto Photon Time | = 2015-08-01T21:11:12 + 7.65 hrs | | 375.4 | 0.01422 |
| Pluto Photon Energy = 312 eV | Pluto Photon Time | = 2015-08-01T21:11:12 + 16.8 hrs | | 376.6 | 0.01096 |
| Pluto Photon Energy = 477 eV | Pluto Photon Time | = 2015-08-01T21:11:12 + 20.7 hrs | | 377.0 | 0.00873 |

[a] - Total On Target Time (TOTT) = 35.1+14.1+17.6+107.2 = 174.0 ksec
[b] – Pluto 0.31 - 0.60 keV individual pointing statistics: 24 Feb 2014: 2.00 - 0.221 ± 0.453 ; 26 Jul 2015: 1.00 - 0.092 ± 0.305 ; 30 Jul 2015: 1.00 - 0.128 ± 0.348; 01 Aug 2015: 4.00 - 0.770 ± 0.969
[c] - Total Net Pluto Counts = 8.0-(0.221+0.092+0.128+0.770) = 6.79; Confidence for 6.79 cts in 174 ksec vs. 1.2 cts background > 5 x 10$^{-4}$
[d] - As determined from New Horizons SWAP measurements within 4 AU upstream (Feb 2014) and 0.2 AU downstream (Jul-Aug 2015) of Pluto

# Figures

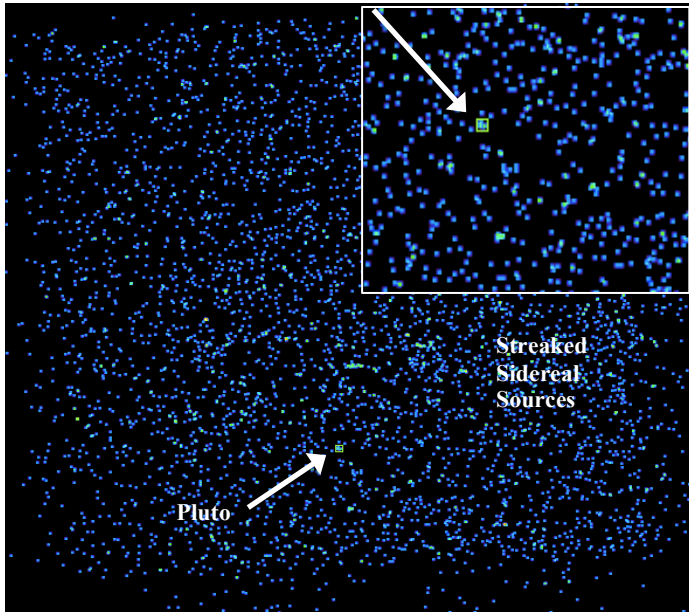

**Figure 1 – Combined results of our *Chandra* ACIS-S 2014-2015 Pluto observations. The 0.31 - 0.6 keV events from all 4 epochs (174 ksec total on-target time) have been co-added in a Plutocentric frame moving with the planet.** Fixed background sources, mostly faint in this passband, are trailed horizontally. The ephemeris position of Pluto is denoted by the red arrow. Inset at upper right: All 8 detected photons lie within an 11 pix x 11 pix box (90% energy) centered on this position, and there is no obvious background source confused with Pluto during the 4 different visits from Feb 2014 through Aug 2015.

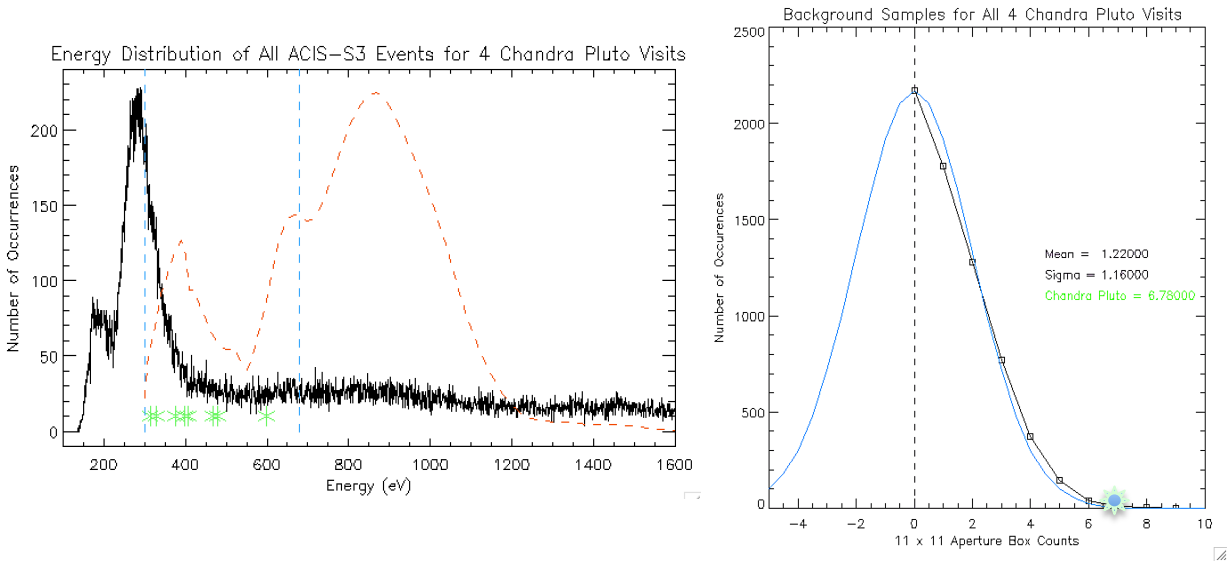

**Figure 2 – *(Left)* Total event spectrum for the *Chandra* Pluto observations.** The spectrum is typical of the background spectra measured for the ACIS-S detector, rising strongly shortwards of 0.35 keV and relatively flat from 0.4 to 1.0 keV. It is unlike the measured 8 count Pluto spectrum (green crosses), which are all clustered in the 0.3 - 0.6 keV region typical of SW CNO minor ion charge exchange (dashed blue line region), with none seen at 0.6 - 1.2 keV especially not near the expected maximum emission created by coherent and incoherent X-ray scattering (dashed red line). *(Right)* **Frequency distribution plot of the number of events in a 11 pixel x 11 pixel box (90% flux footprint)** for 1000 footprints placed randomly across the *Chandra* Pluto-frame combined image. The green star denotes the 6.8 cts found for Pluto after removal of average background value of 1.2 counts. The statistical significance of the detection is evident.

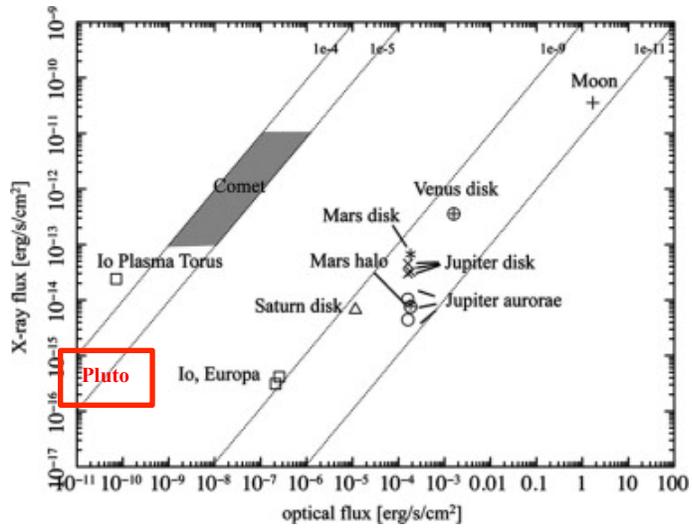

**Figure 3 – X-ray vs. optical flux levels for solar system objects detected in the X-ray.** Pluto's relatively high X-ray/optical flux ratio is similar to that of comets and the Io plasma torus in the Jupiter system. (After Dennerl *et al.* 2012)